\documentclass[10pt, conference]{IEEEtran}
\IEEEoverridecommandlockouts
\usepackage{cite}
\usepackage{amsmath,amssymb,amsfonts}
\usepackage{graphicx,setspace}
\usepackage{textcomp}
\usepackage{xcolor}
\PassOptionsToPackage{hyphens}{url}\usepackage{hyperref}
\usepackage{algorithm,algpseudocode}
\usepackage{textcomp}
\usepackage{array}
\usepackage{tabularx}
\usepackage{booktabs}
\usepackage{mathtools}
\usepackage{comment}
\usepackage{amsthm}
\usepackage{url}

\usepackage{epstopdf}

\usepackage{subfigure}
\usepackage{makecell}
\usepackage{comment}
\usepackage{marvosym}
\newcommand{\makecellL}{\makecell[l]}%
\usepackage{booktabs}
\newcommand{\tabitem}{~~\llap{\textbullet}~~}
\theoremstyle{definition}

\allowdisplaybreaks

\begin{document}
\title{3D Aerial Highway: The Key Enabler of the Retail Industry Transformation 
}

\author{\IEEEauthorblockN{  Nesrine~Cherif, Wael Jaafar, Halim Yanikomeroglu, and Abbas Yongacoglu}
}

\maketitle

\begin{abstract}
The retail industry is  facing an inevitable transformation worldwide, \textcolor{black}{which is accelerating with} the current pandemic situation. Indeed, consumer habits are shifting from brick-and-mortar stores to online shopping. The bottleneck in the online shopping experience remains the efficient and \textcolor{black}{fast} delivery to consumers. In this context, unmanned aerial vehicle (UAV) technology is seen as a potential solution to address cargo delivery issues. Hence, the number of cargo-UAVs is expected to \textcolor{black}{increase} in the next few decades and the airspace to become \textcolor{black}{crowded}. To successfully deploy UAVs for mass cargo delivery, seamless and reliable cellular connectivity for \textcolor{black}{cargo-UAVs} is required. \textcolor{black}{Thus,} organized and ``connected'' routes in the sky \textcolor{black}{are needed}. Like highways for \textcolor{black}{vehicles}, 3D routes in the airspace should be designed to fulfill \textcolor{black}{cargo-UAV} operations safely and efficiently. We refer to these routes as ``3D aerial highways''. In this \textcolor{black}{article}, we investigate the feasibility of the aerial highways paradigm. First, we discuss the \textcolor{black}{related} motivations and concerns. Then, we present our \textcolor{black}{aerial highways paradigm design}. Finally, we present \textcolor{black}{linked} connectivity issues and potential solutions.
\end{abstract}

\section{Introduction}
Unmanned aerial vehicles (UAVs) are gaining momentum in a wide range of applications, such as search-and-rescue, surveillance, and on-demand cellular connectivity \cite{mozaffari2018tutorial}. Specifically, UAVs \textcolor{black}{contribute in solving} the logistics of the delivery industry \cite{peng2019hybrid}. With \textcolor{black}{the proliferation of}
online shopping, an increasing amount of cargo has to be delivered in a timely manner. For instance, Amazon, FedEx, and UPS are delivering approximately 2.5, 3, and 4.7 billion US packages every year, respectively. With already congested roads, \textcolor{black}{delayed} truck deliveries \textcolor{black}{complicate} tight deadlines (https://www.cnbc.com/2019/12/12/analyst-amazon-delivering-nearly-half-its-packages-instead-of-ups-fedex.html).

\textcolor{black}{Alternatively, UAV technology is seen as an eco-friendly and cheaper mean of delivery for light-weight cargo \cite{stolaroff2018energy}.} Indeed, UAVs fly for moderate/long distances \textcolor{black}{collision-free, due to} sophisticated sense-and-avoid techniques. \textcolor{black}{Also, using cargo-UAVs cuts} delivery costs by at least 66\% \cite{sudbury2016cost}. 

Amazon is \textcolor{black}{pioneering development of} UAV-based \textcolor{black}{platforms} for goods delivery in hard-to-reach and remote areas \textcolor{black}{with} ``Amazon Prime Air''. 
\textcolor{black}{Through project ``Skyways'', Airbus} validated the feasibility of end-to-end \textcolor{black}{UAV-based} parcel delivery for shore-to-ship missions dedicated \textcolor{black}{to} enhanced maritime logistics. Moreover, they demonstrated successful cargo delivery in dense urban environments.
Food delivery industry is also seizing the opportunity to modernize operations in cities. For instance, UberEats started testing a UAV-based delivery system over San Diego in 2019. Obviously, the fast integration of UAV technology into different sectors shows that, within few years, UAV-based delivery will be \textcolor{black}{operating} not only in remote areas, but also in dense urban centres.
This will create a \textcolor{black}{high} volume of cargo \textcolor{black}{to deliver} via airspace and in a stringent time frame. Hence, coordinated \textcolor{black}{cargo-UAV operations} is needed to guarantee the safety and fluidity of aerial traffic.

In this context, handful research papers investigated the problem of designing efficient UAV routes. \textcolor{black}{Specifically}, \cite{masatoshi2019} 
proposed a \textcolor{black}{safe} drone highways network that eliminates the risk of UAV accidents, \textcolor{black}{while} a more agile \textcolor{black}{UAV routes} implementation, using evolutionary computing, has been presented in \cite{majd2017}. 
\textcolor{black}{Finally, an} urban logistics airport for UAVs was \textcolor{black}{discussed} in \cite{zeng2019}, where the authors proposed UAV flow control based on graph theory. \textcolor{black}{Yet, to the best of our knowledge, none of the works presented a complete design process of coordinated UAV routes for massive cargo delivery operations.}

Since there are more degrees-of-freedom in planning routes in airspace than on the ground, we define a \textit{3D aerial highway} as a set of aerial routes that cargo-UAVs must follow to fly from one location to another.
For simplicity, the term ``route'' is similar to a highway, street, or avenue in a road network, whereas ``aerial highways'' is similar to the road network itself.      
The ``creation" of 3D aerial highways depends mainly on the cellular connectivity of its routes. 
Indeed, cargo-UAVs require a reliable communication link for the following tasks. First, combined with a global positioning system, \textcolor{black}{it allows for} accurate UAV positioning \textcolor{black}{and} precise landing. Also, command-and-control (C\&C) exchanges are critical for the safety and security of UAVs, as the latter report in real-time any physical or cyber-physical anomaly to the control center. Finally, UAVs can periodically map the area to identify any on-going urban changes that may \textcolor{black}{impact the} highways design.

Several research works shed light on potential network architectures to connect UAVs \cite{azari,nesrineglobecom,nesrine2020globecom}. For instance, \textcolor{black}{relying on} terrestrial networks \textcolor{black}{for} aerial coverage has been extensively investigated, and realistic channel models for cellular ground-to-air channels \textcolor{black}{standardized} by the Third Generation Partnership Project \textcolor{black}{(3GPP) TR 36.777}. Results suggest that terrestrial networks may not be adequate \textcolor{black}{to provide}
ubiquitous connectivity to cargo-UAVs. 
Subsequently, different architectures were proposed, including vertical heterogeneous networks (VHetNets) \cite{nesrineglobecom} and standalone aerial networks enabled by UAV base stations (UAV-BSs) \cite{nesrine2020globecom}. \textcolor{black}{Given the particular characteristics of cargo-UAV operations, it is not yet established which network architecture is the most suitable for providing seamless and ubiquitous communications in 3D aerial highways.}

To ensure the safe operations of large-scale cargo-UAV systems \textcolor{black}{in aerial highways}, a \textcolor{black}{non-negligible amount of} data has to be collected and analyzed, including \textcolor{black}{logs on cargo-UAV missions. The latter can be analyzed by artificial intelligence agents and used to update the aerial highways.}


To fully exploit the potential of UAVs for cargo \textcolor{black}{delivery}, we propose here a \textcolor{black}{design process of} 3D aerial highways. In addition, we \textcolor{black}{discuss} concerns related to the 3D aerial highways paradigm, \textcolor{black}{describe} our vision, and \textcolor{black}{study} the suitability of different cellular networks for cargo-UAV \textcolor{black}{operations}.

\begin{figure}[t]
	\centering
	\includegraphics[width=0.9\linewidth]{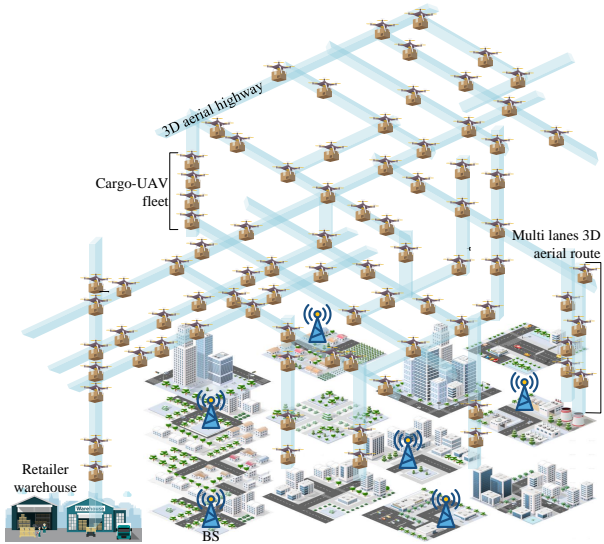}
	\caption{A 3D aerial highway design.}
	\label{fig:sys}
\end{figure}

\section{3D Aerial Highways: Description and Concerns}

\subsection{Description}

Unlike conventional highways, 
3D aerial highways define virtual routes in the airspace. \textcolor{black}{They} can be designed in urban, rural or hard-to-reach areas. For example, Fig. \ref{fig:sys} \textcolor{black}{illustrates} 3D aerial highways above an urban area, where cargo-UAVs travel between the retailer warehouse and consumers. 
\textcolor{black}{These} routes are strategically planned at different altitudes according to specific criteria, e.g., cargo-UAV type, properties, and payload. Moreover, vertical routes are designed for easy transitions and uninterrupted cellular connectivity.

3D aerial highways present attractive characteristics. First, they are flexible and easily reconfigurable. Such qualities are handy in \textcolor{black}{case} of cellular network failures or out-of-control ground/sky layout modifications. 
Also, energy-efficient command-and-control can be achieved using cargo-UAV swarms. When a cargo-UAV fleet is deployed in an area, UAVs heading in the same direction can delegate a cargo-UAV to monitor all C\&C \textcolor{black}{exchanges}, thus reducing the fleet's communications \textcolor{black}{with the control center} to a minimum. 
Finally, aerial highways \textcolor{black}{can support different retailers.}


\subsection{UAV Regulations}
Regulatory authorities developed guidelines for UAV usage \textcolor{black}{that} establish maximum UAV weight and altitude, purpose, and minimum spacing from individuals and sensitive services \cite[Table I]{fotouhiSurvey}. However, most guidelines apply to recreational UAVs \textcolor{black}{only}. Regulations for \textcolor{black}{commercial UAVs should} be more stringent, especially when carrying heavy loads \textcolor{black}{as it} may \textcolor{black}{present} safety \textcolor{black}{risks}. Hence, aerial highways above dense areas require adhering to several restrictions, e.g., no-fly zones, and safe flying above pedestrians and sensitive buildings.

\subsection{Public Safety}
 \textcolor{black}{Unlike conventional aviation operations, UAVs more likely experience failures}. Studies demonstrated that UAV accident rates are very high, due to collisions with structures, aircraft, etc \cite{riskanalysis}. 
 To reduce them, authorities limited the maximum UAV payload (below 30 kg) and flying altitude (below 122 m). When maliciously used, UAVs can trigger public service disruptions. For instance, more than 140,000 travellers were blocked due to UAV sighting at Gatwick Airport, UK \textcolor{black}{in Dec. 2018}. This incident revealed the extent to which \textcolor{black}{UAVs} can endanger daily life. 

\subsection{Privacy and Security}
As UAV technology \textcolor{black}{developed new applications}, 
the privacy of individuals and communities \textcolor{black}{become} under threat. \textcolor{black}{Indeed, cargo-UAVs} equipped with sophisticated sensors and cameras are sensing and collecting data, such as location addresses and aerial photos. This data can be hacked or stored in offshore unsecured data-centres. Retailers, who operate their online deliveries via cargo-UAVs, are responsible for securing collected data and protecting it from cyber-attacks. \textcolor{black}{Thus, the cargo-UAVs manager must deploy the most advanced and secured C\&C exchange protocols, e.g., data encryption and blockchain-based data transmission, to guarantee not only the privacy and security of cargo-UAV data, but also the safety of humans and properties on the ground.}

\subsection{Social Acceptability}
The emergence of UAV-based applications has generated \textcolor{black}{different} responses from the public, \textcolor{black}{which depended} on the use-case. 
Specifically, risk assessment, privacy concerns, and job security impact, are the main factors influencing the social acceptability of the UAV technology. For instance, UAVs \textcolor{black}{are} positively perceived by farmers as they \textcolor{black}{contribute for} food security.
However, their use in urban areas can be unwelcome due to impact on job losses and risk to properties and \textcolor{black}{individuals}.
In a recent survey conducted on the public perception of UAVs \cite{riskanalysis}, the respondents did not overrate the risk and threat of UAVs, compared to manned aircraft. However, privacy issues, military use, and UAV misuses emerged as prevalent concerns. The authors concluded that the perception of UAVs has yet to be formed and, as \textcolor{black}{the} technology matures, its acceptability will evolve positively.

\section{3D Aerial Highway: Our Design Vision}

\begin{figure*}[t]
	\centering
	\includegraphics[width=0.92\linewidth]{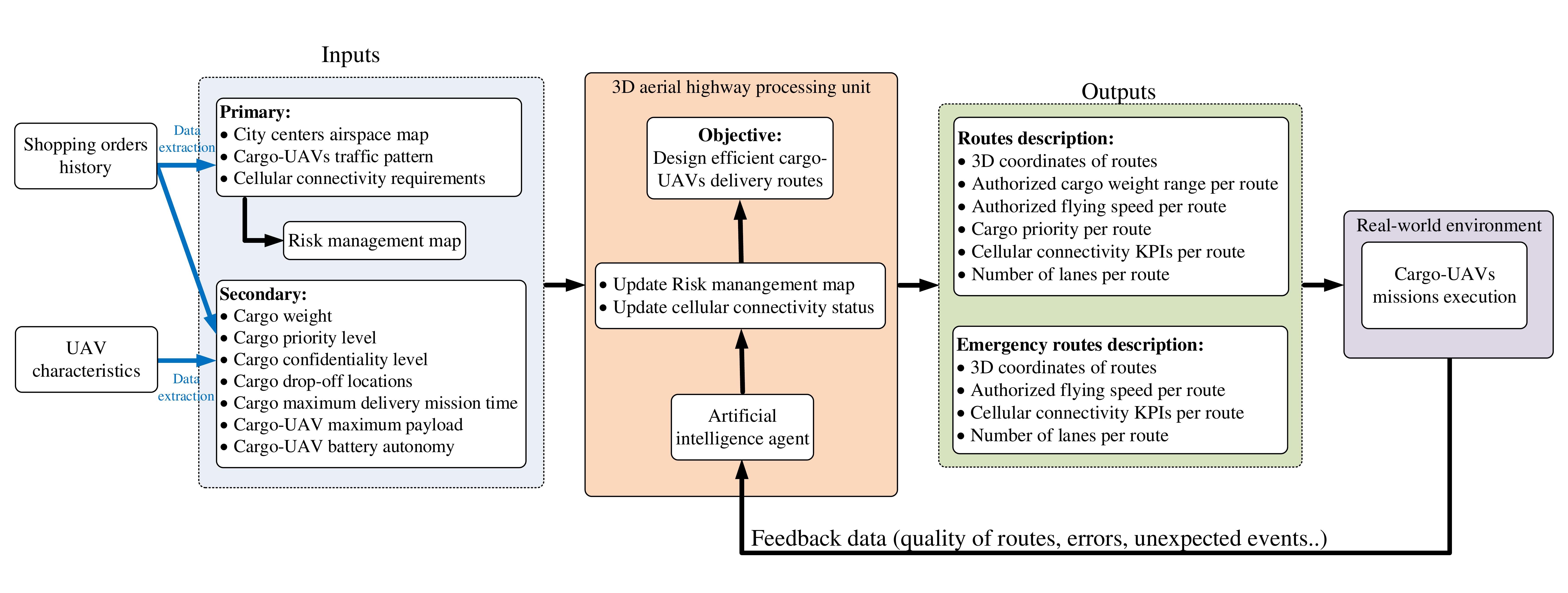}
	\caption{Our vision of the 3D aerial highway design \textcolor{black}{process}.}
	\label{fig:diag}
\end{figure*}

Enabling massive cargo-UAV delivery requires rigorous 3D aerial highway design, where several parameters \textcolor{black}{are considered}. 
\textcolor{black}{This design system sets up the regulatory routes that UAVs must follow when the latter plan their missions. This role can be attributed to a country's regulations entity, such as the Federal Aviation Administration (FAA) in the US. Due to the large sky volume to cover, the design process cannot be handled by a unique central entity. Alternatively, smaller geographical areas can be defined where several design entities can be deployed and each one manages the 3D aerial highways design in its specific geographical area. In such systems, neighbouring design entities exchange data in order to align their highways at the edges of their areas.}


In Fig. \ref{fig:diag}, we depict the envisioned 3D aerial highway design \textcolor{black}{process}, including a description of the input parameters, processing unit, and output metrics.

\begin{figure}[t]
	\centering
	\includegraphics[width=0.9\linewidth]{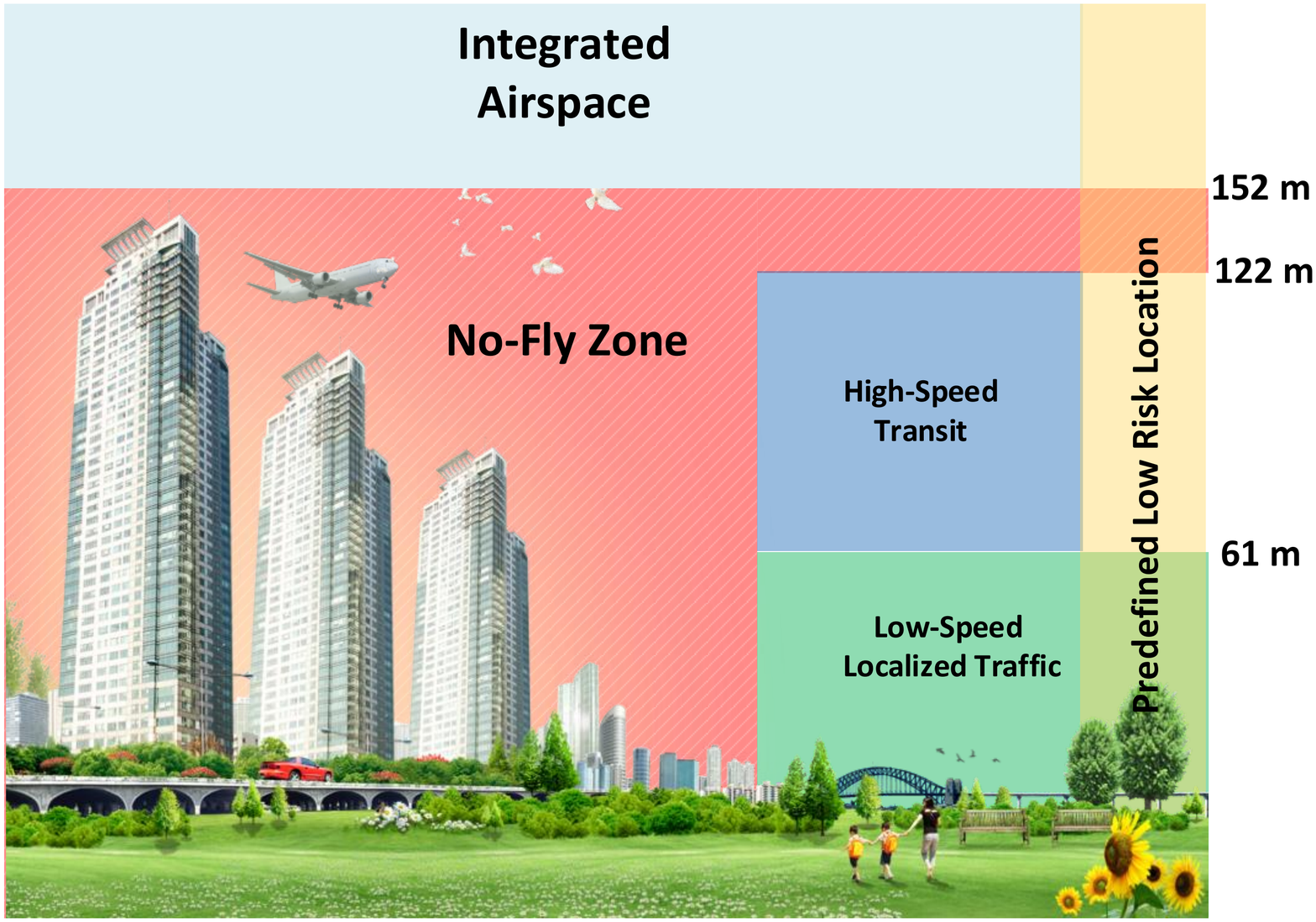}
	\caption{Amazon's airspace segregation model.}
	\label{fig:amazon}
\end{figure}

\subsection{Inputs}
 
\textcolor{black}{The 3D aerial highways design follows a two-step process, which rely on input data of high and low importance. Highly important data, called primary data, drives the selection of the airspace locations to serve as aerial routes, while data of low importance, a.k.a., secondary data, leads the association of the selected aerial routes to different cargo characteristics and UAV capabilities.}

\subsubsection{Primary Inputs} 
They are fundamental in defining the potential locations of aerial highways. They include:

\begin{itemize}

 \item \textbf{City centres airspace map:} \textcolor{black}{It} is a mixture of the urban topography (i.e., buildings, streets) and defined areas of the airspace. These areas can be delimited as suggested by Amazon (Fig. \ref{fig:amazon}). Specifically, cargo-UAVs operating beyond visual line-of-sight \textcolor{black}{(BVLoS)} travel 
 in the ``High-Speed Transit Zone'', while recreational activities \textcolor{black}{occur} in the ``Low-Speed Localized Traffic Zone''. The airspace between 122 m and 152 m is a permanent ``No-Fly Zone'', except for emergencies. Finally, \textcolor{black}{Amazon’s model includes ``Predefined Low Risk Locations'', which are areas with minimal threat to individuals and properties, e.g., wooded areas and deserted fields.}

\item \textbf{Cargo-UAVs traffic pattern:}
From shopping order history, traffic pattern can be extracted, which characterizes the density of order traffic by area. 
\textcolor{black}{Thus}, an area where the number of shopping orders is higher than average should be supplied by a higher number of routes to avoid aerial congestion. 

\item \textbf{Cellular connectivity requirements:} 
    For \textcolor{black}{BVLoS} operations, reliable cellular connectivity would allow C\&C data exchange and cargo-UAV continuous localization, \textcolor{black}{thus preventing} any issue during \textcolor{black}{missions}. Connectivity requirements include \textcolor{black}{mainly} end-to-end communication delay (tens of milliseconds), and tolerable disconnectivity rate of the itinerary, defined as the ratio of \textcolor{black}{disconnected} flight duration to total flight time.
    
\item \textbf{Risk management map:} Cargo-UAVs face several risks during delivery, such as physical attacks by projectiles or birds, or cyber attacks by aerial and ground adversaries. \textcolor{black}{To reduce these risks, a \textit{risk management map} is created and periodically updated based on feedback from cargo-UAVs. This map is designed and then readjusted to address public concerns by allowing cargo-UAVs to travel only in safe and secured routes.}

\end{itemize}
\subsubsection{Secondary Inputs} 
\textcolor{black}{They} are required to organize \textcolor{black}{the potential routes}.
They are composed of:

\begin{itemize}
    \item \textbf{Cargo weight:} Depending on the cargo weight, the latter will be assigned to a specific type of UAV that handles \textcolor{black}{it} and will follow an itinerary composed of routes dedicated to this range of weights.
    
    \item \textbf{Cargo priority level:} Cargo may have different priority levels (e.g., standard, premium, or urgent), causing the delivery to be scheduled differently and/or put on a different priority itinerary. For instance, premium cargo can be delivered in a shorter time by traveling in more direct \textcolor{black}{priority routes}, while urgent cargo \textcolor{black}{have access to routes in the ``No-Fly Zone''.}
    
     \item \textbf{Cargo confidentiality level:} Cargo content may have different levels of confidentiality. Delivering official documents to citizens (e.g., passports, government ID) should be treated with high security measures. As such, UAVs must be equipped with high-end encryption protocols and use the safest aerial routes 
     to guarantee \textcolor{black}{mission integrity}.
     
    \item \textbf{Cargo drop-off locations:}  Aerial routes are expected to link the retailer warehouse to any possible shipping address within the UAV's flying range.
    \textcolor{black}{For neighborhoods with heavy delivery traffic and hard access to private addresses, a common ``drop-off location'' can be established within a walking distance to simplify the delivery process.}

    \item \textbf{Cargo maximum delivery mission time:} 
    \textcolor{black}{According to Amazon, consumers expect delivery within two hours or the same day.}
    Consequently, significant pressure is put on the delivery process, where the maximum cargo-UAV delivery time, \textcolor{black}{since leaving the warehouse until delivery,} becomes crucial to  the end-to-end shopping experience.  

    \item \textbf{Cargo-UAV maximum payload:} 
    Available UAVs for cargo transportation have different payload capabilities, ranging from few hundred grams to hundreds of kilograms. However, \textcolor{black}{current regulations limited the total payload of the UAV, including cargo, to 30 kg}, due to safety concerns of heavy-weight cargo.
    \textcolor{black}{Subsequently, routes} for different ranges of cargo weights, shapes, and solidity have to be defined.

    \item \textbf{Cargo-UAV battery autonomy:} The on-board battery lifetime is a limitation \textcolor{black}{to consider} in the design of aerial routes. Intuitively, \textcolor{black}{more direct routes should be designed for farther drop-off locations.}
    

    \end{itemize}

\subsection{3D Aerial Highway Design Processing Unit}
    
 \textcolor{black}{The processing unit is the core of the 3D aerial routes design system. Its main role is to design and sustain the aerial highways. The process of designing aerial highways undergoes two steps as follows:}
\begin{enumerate}
    \item \textcolor{black}{\textbf{Initial design:} 
    Based on the input data, the processing unit designs 3D aerial highways, which are identified with start and end 3D coordinates. This process is conducted while aiming to maximize a multi-objective function that involves utility functions reflecting the stability of aerial highways in terms of safety, security, connectivity, and taking into account cargo characteristics and UAV capabilities.}
    \item \textcolor{black}{\textbf{Design update:} Since several cargo-UAV related parameters may change over time, such as the city landscape, the shopping patterns, the cellular connectivity, and the security risk, 3D aerial highways must be regularly reconfigured to keep sustaining the cargo-UAV operations efficiently. Practically, unexpected events and changes are fed back to the processing unit via the cargo-UAVs and analyzed using artificial intelligence. Specifically, reinforcement learning algorithms can be leveraged to understand the varying 3D aerial highways environment, optimize their creation and modification over time, and maximize the related multi-objective function.} 
\end{enumerate}

    \subsection{Outputs}
    
    The processing unit produces several metrics, which are summarized as follows:
   
    \begin{itemize}
        \item \textbf{3D coordinates of routes:} As in road networks, aerial routes \textcolor{black}{are} identified mainly by their 3D coordinates. Each route occupies a 3D volume that delimits its boundaries. It can be used for several lanes with smaller 3D volume dimensions. The definition of a route's volume depends on the collision risk level of the area underneath it. 
        
        \item \textbf{Cargo priority per route:} Each aerial route is assigned a priority level that, for convenience, would allow 
        flying cargo with the same priority or higher.
        \item \textbf{Authorized cargo weight range per route:} 
        Each aerial route \textcolor{black}{supports} a range of UAV payloads depending on altitude and regulations. For instance, heavy-payload UAVs travel in low-risk areas, i.e., with minimum ground damage risk for \textcolor{black}{pedestrians} and properties.

         \item \textbf{Authorized flying speed per route:} 
         UAVs may move along aerial highways at different speeds due to their characteristics. To reduce collision risk, routes can be divided for different speed ranges, e.g., \textcolor{black}{fast, moderate, and slow}, within the regulation limits.
         
        \item \textbf{Cellular connectivity KPIs per route:} Each route will be characterized by cellular connectivity KPIs, that are expected to exceed the requirements provided by inputs.
        
         \item \textbf{Number of lanes per route:} Dense neighborhoods may have a \textcolor{black}{large} number of cargo deliveries, thus, aerial routes may need to support \textcolor{black}{this} substantial number. Hence, \textcolor{black}{designing}
           several lanes \textcolor{black}{per} route \textcolor{black}{shortens} the delivery time. To avoid collisions between cargo-UAVs in adjacent lanes, the 3D volume of each lane should be \textcolor{black}{delimited} to the size of cargo-UAVs, while providing enough motion flexibility to avoid dynamic obstacles, e.g., birds and kites, \textcolor{black}{within the} lane's limits. 
           
          \item \textbf{Emergency routes:} For unexpected events, e.g., extended cellular disconnectivity and UAV malfunctioning, the cargo-UAV should rapidly update its itinerary and switch to the reserved emergency route for safe pull-back to a designated ground station. Moreover, due to their robust communication links and high-level safety, with respect to the \textit{Risk management map}, emergency routes can be used to transport critical supplies in case of natural disasters or life-threatening situations.\\   
       
    \end{itemize}

\section{Cellular Connectivity for 3D Aerial Highways: A Close Look}
In this section, we focus on  cellular connectivity of cargo-UAVs in 3D aerial highways. Specifically, we discuss different architectures that guide the design of aerial highways for massive cargo-UAV operations.

\subsection{Existing Terrestrial Network}
Since terrestrial networks were designed to cover terrestrial users, their aerial coverage is unreliable as there are many coverage holes in the sky. Moreover, UAV  may be affected by the strong line-of-sight (LoS) interference from other terrestrial-BSs (3GPP TR 36.777). 

\begin{figure}[t]
	\centering
	\includegraphics[width=0.9\linewidth]{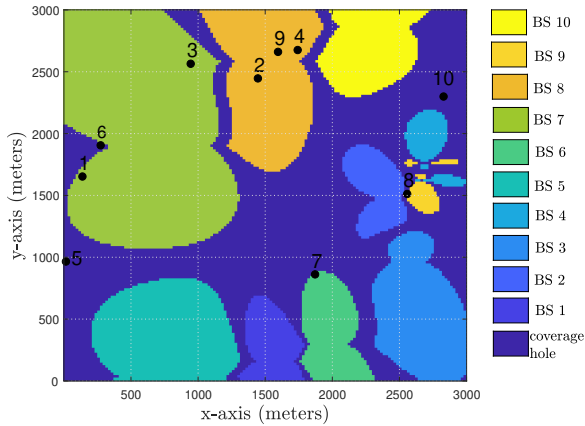}
	\caption{Map of the strongest BS signal per cargo-UAV position. Area of $3000 \times 3000$ m$^2$ in an urban macro (UMa) environment (3GPP 36.777), where 10 terrestrial-BSs with 3D antenna radiation patterns are deployed (3GPP 38.901). Cargo-UAVs fly at 80 meters altitude.}
	\label{fig:aerialcov}
\end{figure}

In Fig. \ref{fig:aerialcov}, we depict the aerial coverage of terrestrial BSs for a cargo-UAV, where the downlink communication is considered successful when the signal-to-interference (\textsf{SIR}) is above a threshold of 5 dB. 
We can see that there are most likely gaps in aerial coverage. 
Moreover, 
we notice that the cargo-UAV is served by BS antenna's side-lobe, e.g., BS 10.

\subsection{Dedicated Terrestrial Network}
As in aviation, where all communications are supported by dedicated terrestrial networks, cargo-UAVs traveling in 3D aerial highway can be served by a similar cellular network design. To this end, terrestrial-BSs with antennas tilted up to the sky can be deployed to cover aerial routes.
However, the coverage of such a dedicated network may face significant challenges when the aerial routes are dynamically reconfigured due to traffic pattern changes or unforeseen events. 
For instance, when a route starts to be congested with a high number of cargo-UAVs, new lanes and/or routes have to be rapidly configured to support this additional traffic load. The cellular connectivity for the new routes has to be guaranteed, which means that terrestrial-BS coverage of the sky has to be potentially reconfigured. Such flexibility may not be available with a dedicated terrestrial network.  Furthermore, this option may not be the most economically attractive due to the need for high capital and operational expenditures.

\subsection{UAV-BS Aerial Network} 
Recently, the use of UAV-BSs to provide cellular connectivity has gained attention since promoted as complementary to terrestrial networks in several use-cases. For instance, when a spike in data rate demand occurs due to a temporary event, e.g., a concert or sporting event, UAV-BSs can be easily deployed to support the extra traffic load.
In cargo-UAV systems, connectivity can also be supported by UAV-BSs. \textcolor{black}{Specifically}, the latter can be placed \textcolor{black}{strategically} along aerial routes to provide connectivity. In such a design, the UAV-BS antenna main-lobes have to be aligned with the cargo-UAV routes.
When new routes are configured, mobility of UAV-BSs allows to move to more adequate locations, thus guaranteeing connectivity for cargo-UAVs along \textcolor{black}{new routes}. 

Although UAV-BSs have extra degrees-of-freedom, i.e., deployment flexibility and mobility, they suffer from limited flying times, 
which complicates their utility. To bypass these constraints, researchers and industry players are investigating several options, including on-demand deployments depending on cargo-UAV traffic, on-the-fly UAV-BS swapping, laser charging, \textcolor{black}{and tethered UAV-BSs that use permanent cables for energy supply and backhauling.}

    \begin{table*}[t]
	\caption{Comparison between different types of cellular networks for cargo-UAV connectivity}
	\label{tab:proscons}
\centering
\small
\begin{tabular}{|l|l|l|}
	\hline
		\bfseries{\makecell[l]{Type of network\\ }}   & 
		\bfseries{\makecell[l]{Pros}}  & \bfseries{\makecell[l]{Cons}}\\ \hline 
		\makecell{Existing terrestrial \\network} & 
		\makecellL{\tabitem \$0 deployment cost\\ 
		\tabitem Mature technology\\
		\tabitem Reliable backhaul links}
		 &  \makecellL{\tabitem Limited cellular aerial coverage\\ 
		\tabitem Fixed deployment of BSs \\
		\tabitem Complex management for aerial \\and ground users} \\ \hline
			\makecell{Dedicated terrestrial \\network} & 
		\makecellL{ 
		\tabitem Strong cellular aerial coverage\\
		\tabitem Mature technology\\
		\tabitem Reliable backhaul links}
		 &  \makecellL{ \tabitem Costly and fixed BS deployment\\
		\tabitem Complex management for dynamic \\3D aerial routes
		} \\ \hline
		\makecell{UAV-BS aerial network}  & 
		\makecellL{\tabitem Easy and quick on-demand deployment\\ 
		\tabitem Reliable cellular coverage for aerial routes \\
		\tabitem Flexible reconfiguration for dynamic 3D aerial routes }
		 &  \makecellL{\tabitem Potentially unstable backhaul links\\ 
		\tabitem Limited on-board processing power\\
		\tabitem Limited flight duration of UAV-BSs\\
		} \\ \hline
		\makecell[l]{LEO satellite network} & 
		\makecellL{\tabitem Ubiquitous cellular coverage\\ 
		\tabitem Potentially supports low-latency links (LEO round trip \\delay between 2.66 ms and 13.33 ms \cite[Table I]{alam2020high})}
		 &  \makecellL{\tabitem Several challenges still unresolved\\ 
		\tabitem Potentially unstable backhaul links} \\ \hline
		\makecell[l]{HAPS aerial network} & 
		\makecellL{\tabitem Quasi-static location and operation for long \\periods of time (up to 6 months)\\ 
		\tabitem Reliable wide cellular coverage\\
		\tabitem Easy location update\\
		\tabitem Potentially supports low-latency links (HAPS round trip \\delay between 0.13 ms and 0.33 ms \cite[Table I]{alam2020high})}
		 &  \makecellL{\tabitem Potentially unstable backhaul links\\
		 \tabitem Limited on-board battery capacity}\\
		 \hline
	\end{tabular}
\end{table*}

\subsection{LEO Satellite Network}
With the evolution of space technology today, the costs of satellite production and deployment are significantly reduced. This has made LEO satellites more attractive for providing ubiquitous and low-latency communications. SpaceX has taken the lead with Starlink project, which aims to deploy thousands of LEO satellites to provide Internet connectivity worldwide. In the context of 3D aerial highways, such a network design would be beneficial, especially in rural and hard-to-reach areas. However, for UAV applications, LEO communications face several challenges that need to be resolved. For instance, satellite pointing loss due to satellite vibrations or imperfect tracking-and-stabilization mechanisms may affect the quality of communications. 
Moreover, LEO user-equipment may be hard to install and operate. Indeed, current LEO user-equipment requires mechanical satellite tracking, which tends to be hard when mounted on an energy-limited and moving UAV. These challenges have yet to be resolved to enable LEO satellite-connected UAVs.

\subsection{HAPS Aerial Network}
An interesting alternative to LEO satellites is a HAPS system, which offers similar performance with fewer constraints. In recent publications, HAPS have been proposed to act as super macro-BSs with large coverage footprints (up to 100 km) \cite{alam2020high}. 
\textcolor{black}{HAPS systems operate in the stratosphere at the typical altitude of 20 km, fueled mainly by solar panels and rechargeable batteries. They can be of different types, such as balloons, blimps, and aircraft \cite[Table IV]{kurt2020vision}.} They can stay aloft at a quasi-stationary location, thus providing significant benefits over LEO satellites to achieve the goal of ubiquitous connectivity.  \textcolor{black}{The deployment of HAPS  was initially planned for rural areas and disaster relief applications. However, the economical viability of HAPS is a main concern for its success. For instance, Google's Loon project was recently shut down due to its risky investment and poor turnover. Nevertheless, its legacy is transferred to a more ambitious project, which is the HAPSMobile project. In industry, several HAPS start-ups are leading the way towards high-speed connectivity from the stratosphere, including Thales Alenia Space and Stratospheric Platforms Limited.
 In the context of 3D aerial highways, HAPS would provide connectivity for a massive number of cargo-UAVs in rural and urban areas, which is expected to generate high income for several HAPS-based applications such as aerial delivery, aerial taxis, and intelligent transportation services.}
Owing to these capabilities, HAPS systems can act as an adequate cellular connectivity platform for cargo-UAVs traveling in 3D aerial highways, since a HAPS can guarantee a reliable wide coverage with relatively low-latency, especially in densely-populated areas where thousands of cargo-UAVs are expected to be flying around daily. Nevertheless, HAPS main concern lays in its limited on-board energy, needed mainly for propulsion and communication. Hence, advancements in battery technologies will enable the full potential of HAPS in connecting massive cargo-UAV systems and support future ``intelligent aerial transportation systems''.

\subsection{Coverage Probability Evaluation}
    \begin{figure}[t]
	\centering
	\includegraphics[width=0.9\linewidth]{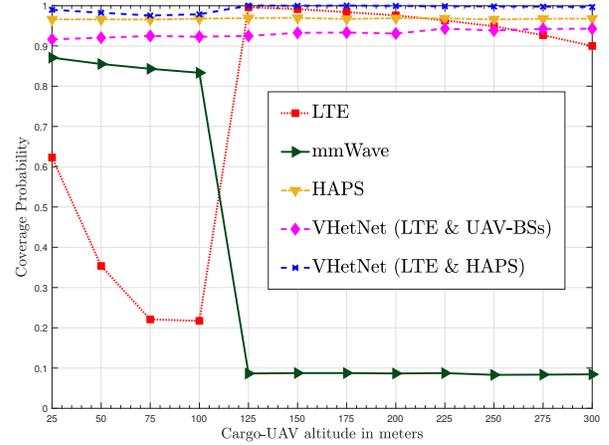}
	\caption{Coverage probability vs. cargo-UAV altitude.}
	\label{fig:casestudy}
\end{figure}

\textcolor{black}{By leveraging tools from stochastic geometry, we present in Fig. 5 the coverage probability performance of a typical cargo-UAV operating in a 3D aerial highway and for different network types, namely the terrestrial LTE and mmWave networks, the HAPS network, and two VHetNets where the first deploys terrestrial LTE BSs and UAV-BSs, while the second uses terrestrial LTE BSs and HAPS. First, the LTE network
provides poor coverage at low altitudes. This is mainly due to blockages such as highrise buildings and trees.
Starting from altitude 125 m, the coverage probability improves as the communication exhibits a higher LoS. 
In contrast, the mmWave network, which operates at frequency 38 GHz and leverages 3D beamforming, behaves inversely to the LTE network, i.e., it demonstrates strong performances at low altitudes due to the low path loss and high antenna gains, but exhibits a low coverage probability at high altitudes caused mainly by a higher path loss impact with distance at high frequencies \cite{rappaport2015}.
Using HAPS, the cargo-UAV enjoys a ubiquitous coverage at any altitude due to the HAPS inherent characteristics providing a wide coverage range of dozens to hundreds of kilometers around \cite{kurt2020vision}. 
Since UAV-BSs complement the terrestrial BSs' coverage, VHetNet (LTE \& UAV-BSs) achieves acceptable performance, which is close to that of HAPS. Finally, VHetNet (LTE \& HAPS) outperforms all network types by providing ubiquitous and reliable coverage at any altitude. Specifically, the latter favors HAPS coverage at low altitudes and LTE coverage at high altitudes, as this strategy guarantees strong LoS communication links. 
Obviously, if cargo-UAVs have to rely on a single network type for connectivity, the HAPS would be the most adequate choice, while full coverage probability is achieved thorough VHetNet (LTE \& HAPS).}

In summary, each of these networks has its advantages and drawbacks in providing cellular connectivity for the 3D aerial highway paradigm. Nevertheless, we envision that a practical deployment for cargo-UAVs will be supported by at least two different types of networks, which will provide reliable connectivity and safe operation in the airspace. 
Specifically, routes can be designed with a prior knowledge of the coverage edges and locations for intra-network and inter-network handovers. Subsequently, a cellular-connectivity strategy can be pre-designed. As the cargo-UAV's trajectory may change during the mission, unexpected disconnectivity events trigger on-board network search to lock on an available network and receive updates for its connectivity strategy. \textcolor{black}{When extended disconnectivity occurs}, it must follow the emergency route for the time to reestablish connectivity and update its strategy.
For the sake of clarity, we provide the pros and cons of these solutions in Table \ref{tab:proscons}.

\section{Conclusion}
In this article, we presented our vision of a 3D aerial highway paradigm, which will be the main enabler of the retail industry transformation. First, we highlighted its main motivations and concerns. Then, we detailed our 3D aerial highway \textcolor{black}{design process} that enables the coordinated and dynamic planning of routes for cargo-UAVs. Finally, we discussed the related issue of cellular connectivity and evaluated possible solutions. For 3D aerial highways to operate safely and effectively, we recommend supporting cargo-UAVs with at least two types of wireless networks.
 \section*{Acknowledgements}
{This work is funded by Huawei Canada and NSERC. The authors thank Dr. Gamini Senarath for insightful discussions.}

\bibliographystyle{IEEEtran}  
\bibliography{references}

\section*{Biographies}
\small{
\noindent \textbf{Nesrine Cherif} [S] (ncher082@uottawa.ca) is a PhD student at uOttawa. Her research interests include non-terrestrial networks.

\noindent \textbf{Wael Jaafar} [SM] (waeljaafar@sce.carleton.ca) is an NSERC Postdoctoral Fellow at Carleton University. His research interests include wireless communications and machine learning.

\noindent \textbf{Halim Yanikomeroglu} [F] (halim@sce.carleton.ca) is a professor at Carleton University, Canada. His research interests cover many aspects of 5G/5G+ wireless networks.

\noindent \textbf{Abbas Yongacoglu} [LM] (yongac@uottawa.ca) is Emeritus Professor at uOttawa. His research area is wireless communications.
}

\end{document}